ORIGINAL RESEARCH

# Proposed DBMS for OTT platforms in line with new age requirements

Aryan Shah • Charmi Shah • Devansh Shah
aryanshah1902@gmail.com   charmishah2611@gmail.com   dev4901@outlook.com

Khushi Shah • Mustafa Africawala • Rushabh Shah
khps2002@gmail.com   mustafa.ace19@sot.pdpu.ac.in   rushabhrs.04@gmail.com

Nishant Doshi
nishant.doshi@sot.pdpu.ac.in

**Abstract** Database management has become an enormous tool for on-demand content distribution services, proffering required information and providing custom services to the user. Also plays a major role for the platforms to manage their data in such a way that data redundancy is minimized. This paper emphasizes improving the user experience for the platform by efficiently managing data. Keeping in mind all the new age requirements, especially after COVID-19 the sudden surge in subscription has led the stakeholders to try new things to lead the OTT market. Collection of shows being the root of the tree here, this paper improvises the currently existing branches via various tables and suggests some new features on how the data collected can be utilized for introducing new and much-required query results for the consumer.

# Table of Contents

# 1 Introduction

Unarguably, 2020 has been a year of pandemic, which led to many businesses shutting down and also giving a major rise to entrepreneurship opportunities throughout the world. The sectors which primarily provided services over the internet or the sectors which allowed the consumer to enjoy the services sitting at home saw a huge rise in their businesses. The obvious reason being the COVID-19 pandemic, one of the sectors which were already on a rise till the first two quarters of 2020- "Over-the-Top" (OTT) platforms, saw exponential growth in the number of subscriptions every month. As per the reports of Boston Consulting Group (BCG), the subscriptions grew by over 60 percent. The trend is expected to stay as life returns to normalcy, said BCG, pointing to the propensity of Indian consumers to now pay for content they are watching. OTT platforms, in particular, are investing heavily in content creation and acquisition as they eye a large member base in India. Already, the number of hours spent per day on digital video in India, according to BCG, has risen by 14.5 percent in the last two years.

As a result, popular OTT service providers such as YouTube, Netflix have seen an instrumental role in the growth of data streaming, recording a staggering 140% rise in video streaming apps in Australia, India, Indonesia, South Korea, and Thailand (App Annie, The state of the mobile 2019). These statistics show that there exists a strong opportunity for OTT service providers to capitalize on digital media as a strong communication channel.

As a result, the prime causes of quality-Competition comes into the party, as competitions increase the platforms tend to increase the features they offer to the consumer, the foremost of them being the user experience that sticks the user to their platform. Due to this many consumers constantly search for products that offer them better features at a cheaper rate. In this paper, we have introduced some of the features to make the user experience better and more efficient ways to manage the data for OTT platforms to equip them to handle queries better.

## 1.1 Our Contributions

The rise in the number of users using OTT platforms in the World especially during the pandemic has encouraged these platforms to bring and adopt new technologies for a better understanding of their users. Media convergence is transforming the media business model. Film and television consumption over the internet force the breakdown of the traditional value chain [4]. In this paper, we have inoculated few unique features that are part of our OTT platform DBMS project and the same are enlisted below, for understanding our users in a better way and providing them what they exactly need

- Here the user is not only seen as a data generator but also we acquire data from the website itself, protecting privacy and encryption. As far the viewing hours are concerned, we filter it according to the shows which are further classified into movies and shows, and with SQL it can be used to generate more complex queries too.
- While our system provides convenience even to new subscribers by using different combinations of relevance shows, related shows and genres in SQL, which not only helps us to collect data in these sectors for our new client but also enhance their experience on our platform apart from others.

- Our data sources include users as well as various websites and the goal of our system is to provide as detailed information about the movies and series as possible which can effectively work even with a small amount of user information.
- In addition to this, our project has the potential to sort the data according to genre, actor name, even actor's age while the particular movie was released, Oscar nominations, PG rating, Inspiration behind the movie and director as well as products related to that movie, and even SQL combinations for this can be applied at the same time.
- Also, we fetch data directly from the respective OTT server and provide the data to the user to reduce data redundancy and provide the following data for each show including series too.
- From the beginning, its prime focus has been on online streaming platforms such as Netflix, Amazon Prime, etc. And the primary goal is to provide as much detailed information as possible right from the created database.

## 1.2 Paper organization

Section 2 gives the literature survey of already existing research papers, types of users of our database and the SQL queries along with output screenshots for the determined models using the data grip platform. In section 3, we give the proposed system of our project consisting of the framework, Entity-Relationship Model and Relational Model. In section 4, we have given the experimental analysis of our project describing the software we used and some sample queries. Finally, in section 5, we conclude and give the future scope of our work. References are given at the end.

## 2 Literature Survey

Since the covid-19 pandemic, there has been an increase in the audience of these OTT platforms making it a very competitive industry. These platforms put in their best efforts to collaborate with high-quality multimedia content that is based on many factors, like the directors, actors, writers, production houses, ratings, Academy nominations, etc. Due to this competition, the content becomes scattered among a range of OTT platforms making it a tedious task for the viewers to watch the content they prefer. A viewer needs to buy, manage and search multiple platforms to watch that single movie or that show.

Multiple papers and models are published that give us a broad idea of how the audience of OTT platforms has grown over time and especially since the pandemic. Some papers have:

(i)  Designed models to manage multimedia content,
(ii) Studied business models of different OTT platforms like Netflix, Hulu, etc.
(iii)That presents different views on how metadata can be modeled, classified, extracted, managed, and applied, to support convenient handling of digital media.[8]

Film and television consumption over the internet force the breakdown of the traditional value chain. [4]

The audiovisual, on-demand, distribution industry is aware that its competitive advantage comes from obtaining information from its users. Thus, Netflix has become the exemplar of internal data management and the use of metadata, demonstrating the use of information that is flexible and adaptable to its environment.[2]

So, we felt the need to develop a model that can relieve the audience from the hassle of searching their favorite shows and movies on various platforms and many more.

## 2.1 Identification of the Different Types of Users and Queries

Our database is directed to three different types of users. These users have access to different information according to the category they fall in.

The three different types of users of our database are:

1. Clients of the database – The people who will use the database for finding information about movies. This group can be further divided into (a) General clients who would search the database for any basic information about the movies, their actors, director, awards won, etc and (b) specialized clients with more complicated search requests like finding all the movies written by a particular writer, finding movies that are inspired by real life incidents, etc.

2. Contributors to the database – The people who will be adding new information to the database. These people are different from the system administrators because they can only add information based on some prespecified constraints.

3. System administrators -These people manage, upgrade, alter and program the database. They are the same as database administrators.

## 2.2 SQL Queries

1.The following SQL query lists the number of actors in each country, sorted high to low:

```
SELECT COUNT(Actor_id), Nationality
FROM Actors
GROUP BY Nationality
ORDER BY COUNT(Actor_id) DESC;
```

Output

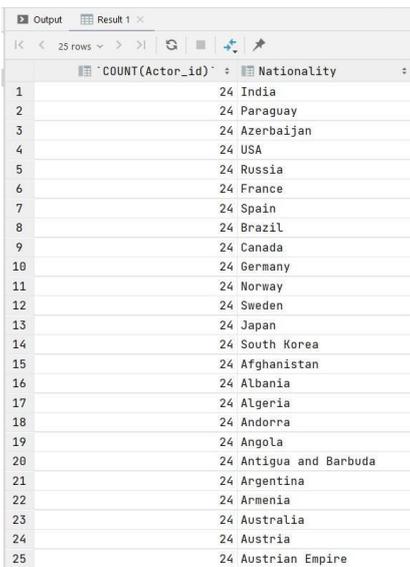

2.The following SQL query lists the shows with IMDB rating= 10

```
SELECT b.`show name`, a.`IMDB rating`
FROM `Critics_Rating` a
JOIN `Show_id-name` b
ON a.show_id = b.show_id
WHERE `IMDB rating` = 10;
```

Output

| # | show name | IMDB rating |
|---|---|---|
| 1 | Make Room for Granddaddy | 10 |
| 2 | Dinah's Place | 10 |
| 3 | From a Bird's Eye View | 10 |
| 4 | The Starlost | 10 |
| 5 | Amy Prentiss | 10 |
| 6 | My Son Reuben | 10 |
| 7 | Adams of Eagle Lake | 10 |
| 8 | Eigener Herd ist Goldes wert | 10 |
| 9 | Kate McShane | 10 |
| 10 | Three for the Road | 10 |
| 11 | Star Maidens | 10 |
| 12 | The Betty White Show | 10 |
| 13 | The Marilyn McCoo and Billy Davis, Jr. Show | 10 |
| 14 | Hedebyborna | 10 |
| 15 | Send in the Girls | 10 |
| 16 | The Lazarus Syndrome | 10 |
| 17 | A Man Called Sloane | 10 |
| 18 | Sapphire & Steel | 10 |

3.The following SQL query lists the show id, show Name and Writer for Shows whose writer is S.S.Wilson.

*SELECT a.`Show Name`, b. Writer, b.`Release year`*
*FROM `Show_id-name` a*
*JOIN `Collections_of_shows` b*
*ON a.Show_id = b.Show_id*
*WHERE b.Writer= 'S.S. Wilson'*

Output

| # | Show Name | Writer | Release year |
|---|---|---|---|
| 1 | For the Love of Ada | S.S. Wilson | 1974 |
| 2 | The Associates | S.S. Wilson | 1988 |

4. If you want to know the views/month for each platform combined, then the group by query would be as follows.

*SELECT b.`Platform name`, SUM(`views/mo`) AS 'TOTAL'*
*FROM Statistics a*
*JOIN Platforms b*
*ON a.Platform_id = b.Platform_id*
*GROUP BY b.`Platform name`;*

Output

| | `Platform name` | TOTAL |
|---|---|---|
| 1 | Eros Now | 23053 |
| 2 | TVF Play | 20637 |
| 3 | ALT Balaji | 23973 |
| 4 | Sony LIV | 21424 |
| 5 | Netflix | 23494 |
| 6 | Amazon Prime VIdeo | 22753 |
| 7 | Hungama Play | 22927 |
| 8 | Disney + Hotstar | 21341 |
| 9 | Voot | 22819 |
| 10 | Jio Cinema | 25369 |
| 11 | Ullu App | 21694 |
| 12 | Mx player | 23371 |

5. The following SQL query will list the shows having the total number of episodes less than 6 and number of seasons less than 2 along with the name of the production house that is producing the show.

*SELECT b.Show_id,a.`Show Name`,c.Production_Name ,b.Seasons,b.Episodes*
*FROM `Show_id-name` a*
*JOIN `TV_series` b*
*ON a.Show_id = b.Show_id*
*JOIN Productions c*
*ON b.Production_id = c.Production_id*
*WHERE Seasons<2*
*AND Episodes <6*
*ORDER BY b.Seasons;*

Output

| | `Show Name` | Production_Name | Seasons | Episodes |
|---|---|---|---|---|
| 1 | Three Men of the City | Forward Media | 1 | 5 |
| 2 | The Georgian House | Orion Pictures | 1 | 5 |
| 3 | Secret Army | Warped Films | 1 | 5 |
| 4 | How the West Was Won | Columbia Pictures Corporation | 1 | 5 |
| 5 | Katitzi | ABC Motion Pictures | 1 | 5 |
| 6 | Kingswood Country | Half Moon Entertainment | 1 | 5 |
| 7 | A Man Called Sloane | Warped Films | 1 | 5 |

6. The following SQL query will list the movies starring a male actor having age less than equal to 40 , having genre Adventure and PG rating of U/A.

```sql
SELECT a.`Show Name`,b.Writer, b.`Release year`, b.Genre,e.`Actor name`
FROM `Show_id-name` a
JOIN `Collections_of_shows` b ON a.Show_id = b.Show_id
JOIN Director c ON a.Show_id = c.Show_id
JOIN `Actor_id-Show_id` d ON a.Show_id = d.Show_id
JOIN Actors e ON d.Actor_id = e.Actor_id
JOIN PG_Rating f ON a.Show_id = f.Show_id
WHERE Age <= 40
AND Genre = 'Adventure'
AND `U/A` = 1
AND Gender = 'Male'
ORDER BY a.Show_id
```

Output

| | `Show Name` | Writer | `Release year` | Genre | `Actor name` | Age |
|---|---|---|---|---|---|---|
| 1 | Toma | Manya Starr | 1985 | Adventure | Henry Mancini | 15 |
| 2 | Harry O | Larry Cohen | 1979 | Adventure | Steve Guttenberg | 14 |
| 3 | Partridge Family 2200 AD | Joe Eszterhas | 2015 | Adventure | Noah Hathaway | 38 |
| 4 | The Sweeney | Joe Camp | 2021 | Adventure | Lance Henriksen | 21 |
| 5 | The Lost Islands | Stewart Raffill | 1980 | Adventure | Diane Keaton | 40 |
| 6 | We'll Get By | David Saperstein | 1985 | Adventure | Heather Langenkamp | 17 |
| 7 | Eight Is Enough | David Webb Peoples | 2000 | Adventure | Kelly McGillis | 28 |
| 8 | En by i provinsen | Stuart Gordon | 2001 | Adventure | Kelly McGillis | 28 |
| 9 | Grange Hill | Terry Rossio | 1979 | Adventure | Colm Meaney | 17 |

# 3 Proposed System

## 3.1 Framework

In our proposed system, the table "Collection of shows" inherits each table from our database.
Various other entities are connected to the main table which states various properties a show can constitute of. All the entities which are connected to the main table contain a foreign key as 'Show_id'. There are several other entities that are interconnected but not connected with the main table. For example, critics rating is connected directly to the collection of shows table as ratings will be purely for the shows and for which show_id is required for the data generation thus it is directly connected to the main table. Whereas the resolution table is connected to the subscription table because the resolution property is based on the platform on which the show is available.

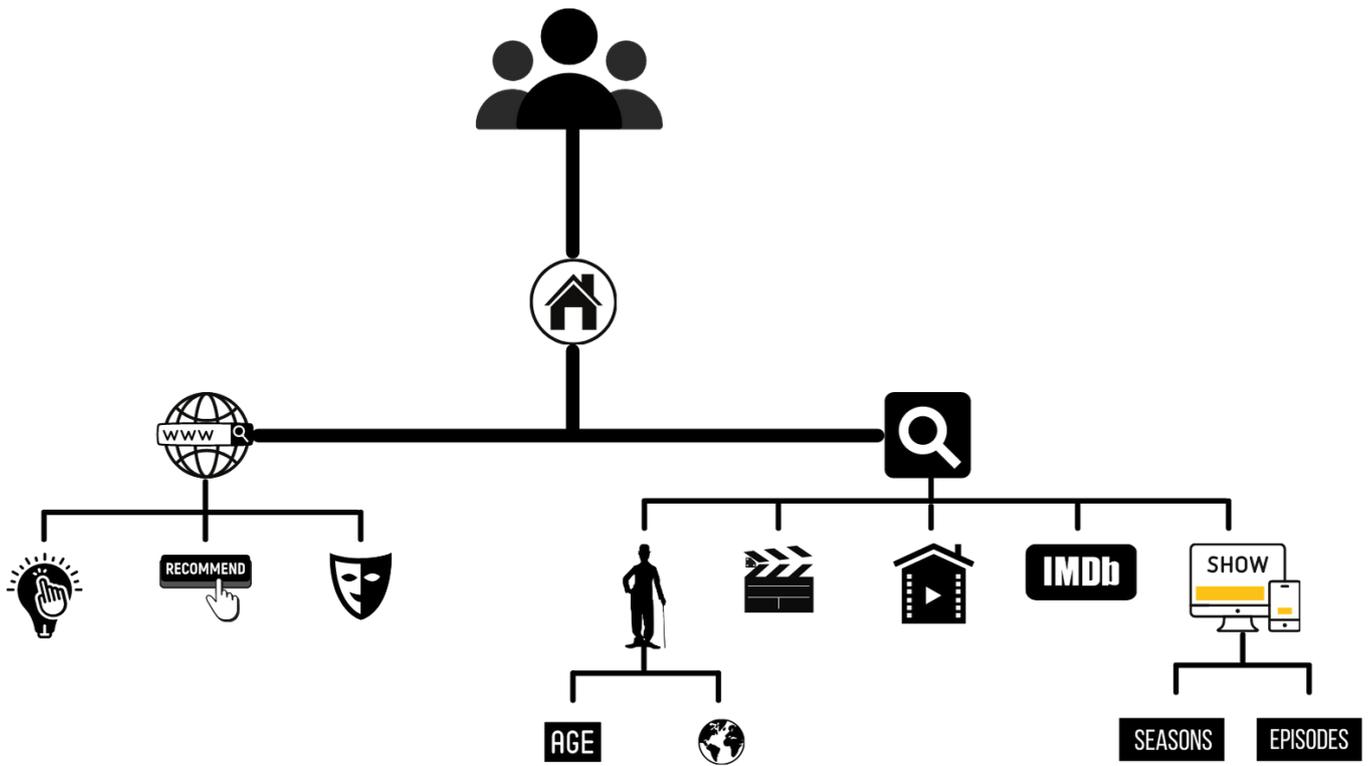

Figure 1. Proposed User Flow

## 3.2 Relational Model

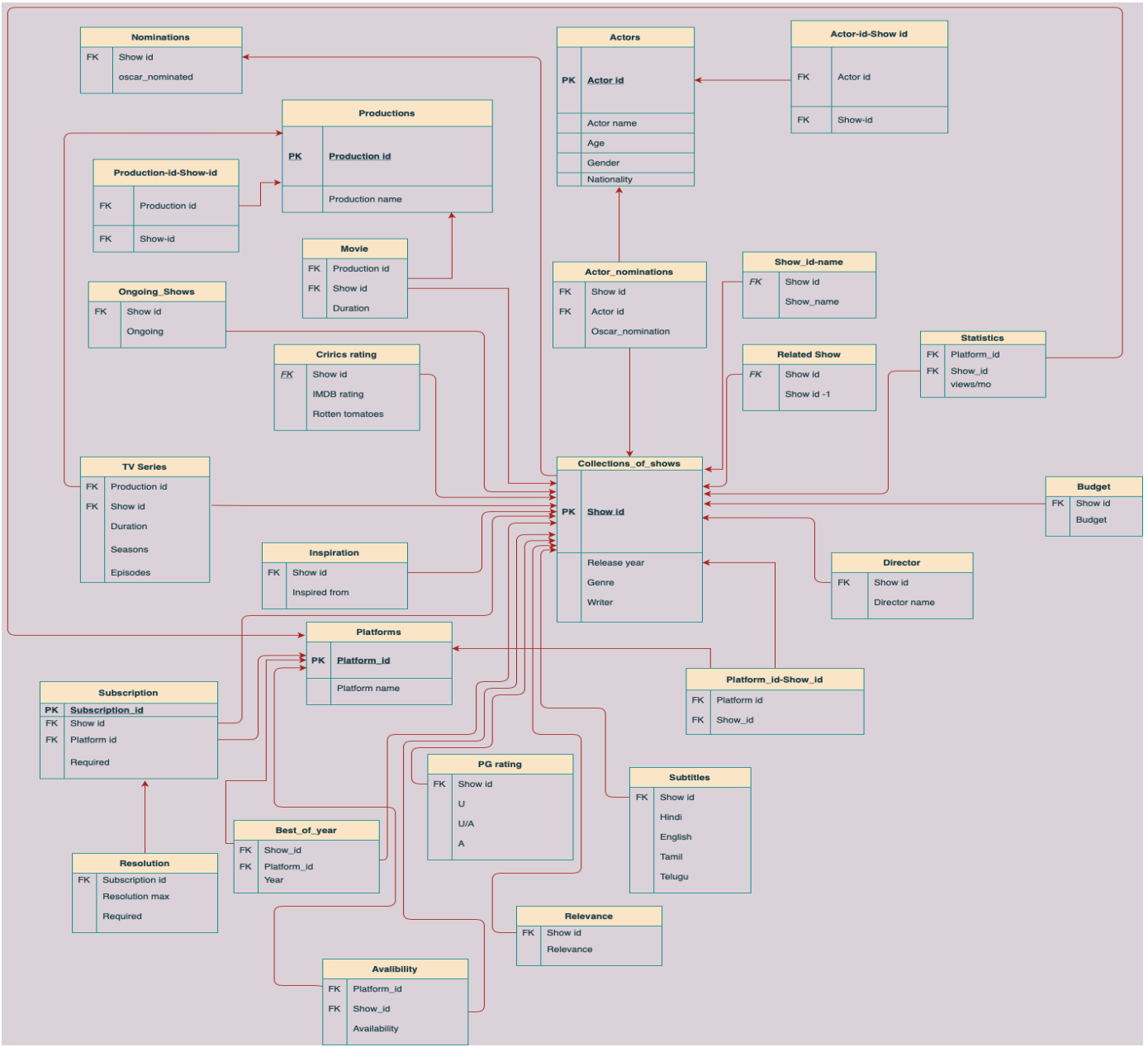

## 3.3 Entity-Relationship Model

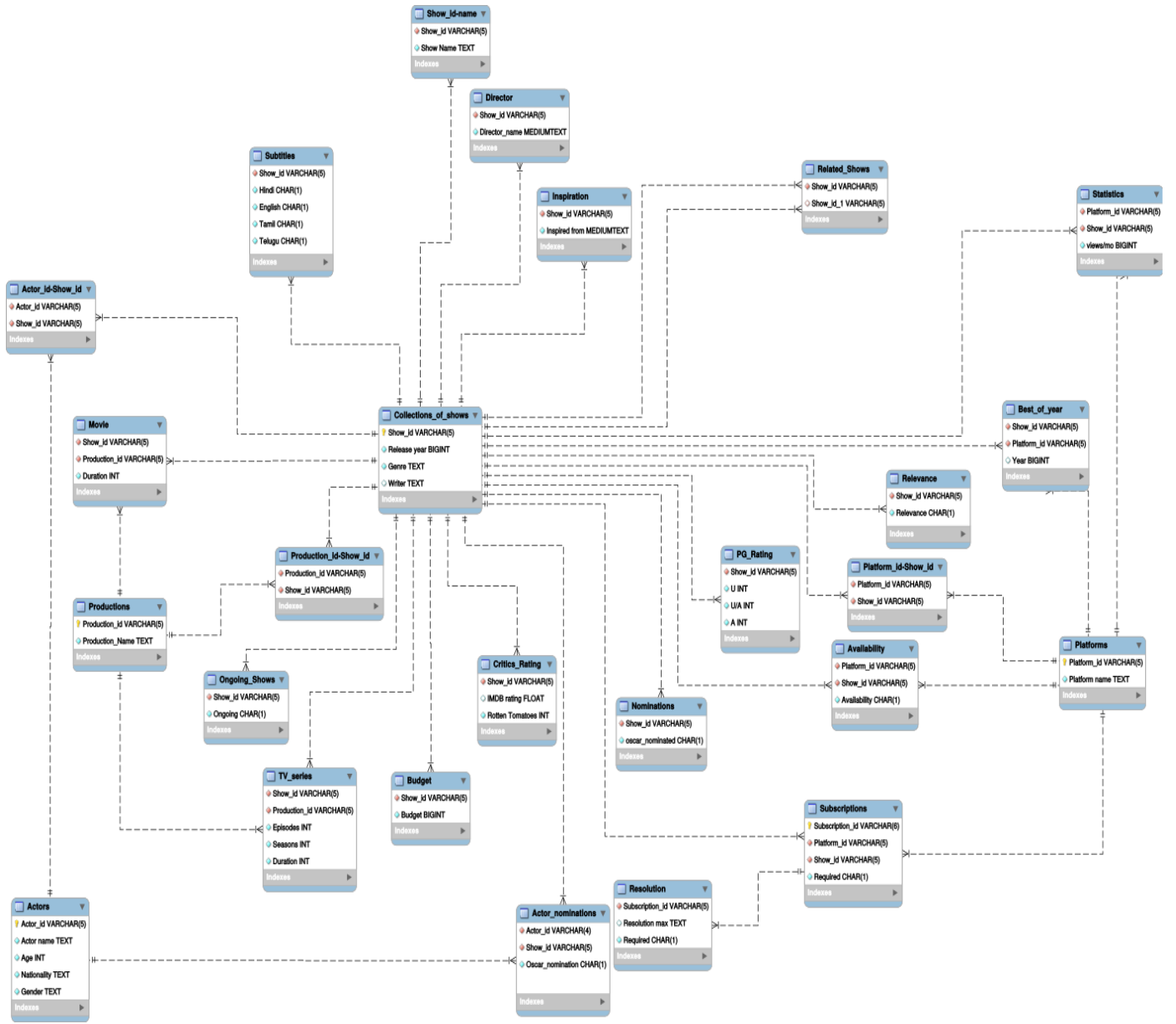

## 3.4  Explaining Schemas

| Schema Construct | Construct Description |
|---|---|
| **Collection of shows** | **Entity class, to model all the shows** |
| • Release year | Year in which the show is released |
| • Writer | Person who wrote the show |
| • Genre | Genre of the show |
| **Show-id-name** | **Entity class, to model name of the show** |
| • show name | Name of the show |
| **Actor** | **Entity class, to model actors involved in the shows** |
| • Actor name | Name of the actor who has worked in the show |
| • Gender | Gender of actor |
| • Age | Age of actor |
| • Nationality | Nationality of actor |

| Schema Construct | Construct Description |
|---|---|
| **Actor-id-Show-id** | **Entity class, to model the connectivity between actor and show** |
| **Production-id-Show-id** | **Entity class, to model the connectivity between production house and show** |
| **Production** | **Entity class, to model the name of production house** |
| • Production name | Name of the production house |
| **Critics rating** | **Entity class, to model rating of the show** |
| • imdb | IMDB rating of the shows |
| • rotten tomatoes | Rotten Tomatoes of the shows |
| **PG rating** | **Entity class, to model PG rating of the show** |
| • U | suitable for age 4 or above |
| • U/A | suitable for age 12 or above |
| • A | suitable for age 18 or above |

| Schema Construct | Construct Description |
|---|---|
| **Platform-id-Show-id** | **Entity class, to model the connectivity between platform and show** |
| **Platforms** | **Entity class, to model the name of platform** |
| • platform name | Name of the platform on which the show is streaming |
| **Subscriptions** | **Entity class, to model whether subscription is required or not** |
| • required(y/n) | Whether subscription is required or not |
| **Availability** | **Entity class. to model availability of the show on platform** |
| • Availability | Whether show is available on a certain platform or not |
| **Relevance** | **Entity class. to model the relevance of show** |
| • Relevance | Whether the show is popular currently or not |
| **Duration** | **Entity class, to model the duration of show** |
| • duration | Duration of the show |
| **Resolution** | **Entity class, to model the resolution of the show** |
| • resolution | Max resolution available for the show |
| • required | Subscription required or not |
| **TV series** | **Entity class, to model the information about the TV show** |
| • duration | duration of the show |
| • season | no of seasons of show |
| • episodes | episodes per season |
| **Subtitles** | **Entity class, to model in which language the subtitles are available** |
| • hindi | available in Hindi |
| • english | available in English |
| • tamil | available in Tamil |
| • telugu | available in Telegu |
| **Ongoing** | **Entity class, to model whether the show is ongoing or not** |
| • ongoing | whether the show is ongoing or not |
| **Director** | **Entity class, to model who is the director of the show** |
| • director | name of the director |
| **Related shows** | **Entity class, to model the shows which are related to each other** |
| **Inspiration** | **Entity class, to model from where the shows are inspired** |
| • inspired from | from where the show is inspired |
| **Nominations** | **Entity class, to model whether the show is nominated for Oscar or not** |
| • oscar nominated(y/n) | whether the show is nominated or not |
| **Budget** | **Entity class, to model the budget of the show** |
| • budget | budget of the show |
| **Statistics** | **Entity class, to model the views/month of the shows** |
| • views/month | views/month of the shows |
| **Best of year** | **Entity class, to model the best show of the year** |
| • year | in which the show of was awarded best of the year |
| **Actor nomination** | **Entity class, to model whether the actor is oscar nominated or not** |
| • Actor Oscar nominated(y/n) | whether the actor is nominated for Oscar or not |

# 4 Experimental Analysis

| Research papers | F1 | F2 | F3 | F4 | F5 | F6 |
|---|---|---|---|---|---|---|
| **Our paper** | ✓ | ✓ | ✓ | ✓ | ✓ | ✓ |
| [2]  | ✓ | ✗ | ✗ | ✗ | ✓ | ✗ |
| [3]  | ✗ | ✗ | ✓ | ✗ | ✗ | ✗ |
| [4]  | ✗ | ✗ | ✗ | ✓ | ✗ | ✓ |
| [5]  | ✓ | ✓ | ✗ | ✗ | ✓ | ✓ |
| [9]  | ✗ | ✗ | ✗ | ✓ | ✗ | ✗ |
| [12] | ✗ | ✓ | ✗ | ✗ | ✗ | ✗ |
| [16] | ✗ | ✓ | ✗ | ✗ | ✗ | ✗ |
| [17] | ✗ | ✗ | ✗ | ✗ | ✗ | ✓ |
| [18] | ✗ | ✗ | ✗ | ✓ | ✗ | ✗ |
| [20] | ✓ | ✓ | ✗ | ✗ | ✓ | ✓ |
| [21] | ✗ | ✗ | ✗ | ✗ | ✓ | ✗ |

| Features | Description |
|---|---|
| F1 | Here the user is not only seen as a data generator but also, we acquire data from the website itself, protecting privacy and encryption. As far as the viewing hours are concerned, we filter it according to the shows which are further classified into movies and shows (and with SQL it can be used to generate more complex queries too.) |
| F2 | While our system provides convenience even to new subscribers by |
|    | using different combinations of relevance shows, related shows and genres in SQL, which not only helps us to collect data in these sectors for our new client but also enhance their experience on our platform apart from others. |
| F3 | Our data sources include users as well as various websites and the goal of our system is to provide as detailed information about the movies and series as possible which can effectively work even with a small amount of user information |
| F4 | In addition to this, our project has the potential to sort the data according to genre, actor name, even actor's age when a particular movie was released, oscar nominations, PG rating, Inspiration behind the movie and director as well as production related to that movie, and even using SQL combinations for this can be applied at the same time. |
| F5 | Also, we fetch data directly from the respective OTT server and provide the data to the user to reduce data redundancy and also provide the following data for each show including series too. |
| F6 | From the beginning itself prime focus has been on online streaming platforms such as Netflix, Amazon Prime etc. And the primary goal is to provide as much detailed information as possible right from the created database. |

## 5 Conclusion and Future Work

The increase in usage of OTT platforms and other online multimedia services has called for an improved database management system. With the rise seen in subscription to various OTT service providers, a fight over quality of content and better user experience has emerged. One of the ways of achieving this is better management of data. By also providing more information to the user on the content and providing improved filters, the same can be achieved. This paper provides information on how a user can be given better experience with the suggested database management system. It can improve searching and sorting of data according to genre, actor name, Oscar nominations, PG rating, Inspiration behind the movie and director as well as production related to that movie. While researchers are always finding ways to collect and use the data effectively, this research aims to contribute to the same and hope that it helps in making better user experiences and better products.

**References**

1. Javagar,M.(2020) .*Tv Shows on OTT Platforms* (CC0:Public Domain)[dataset ] (https://www.kaggle.com/javagarm/tv-shows-on-ott-platforms)

2. Fernández-Manzano, E. P., Neira, E., & Clares-Gavilán, J. (2016). *Data management in audiovisual business: Netflix as a case study. Profesional de La Informacion*, *25*(4), 568–576. https://doi.org/10.3145/epi.2016.jul.06 (Fernández-Manzano et al., 2016)

3.  Clares Gavilán, J. (2014). *Estructura y políticas públicas ante los nuevos retos de la distribución y consumo digital de contenido audiovisual. Los proyectos de Vídeo bajo Demanda de cine filmin y Universciné como estudio de caso. TDX (Tesis Doctorals En Xarxa)*.
    http://www.tesisenred.net/handle/10803/247706
    (Clares Gavilán, 2014)

4.  Izquierdo-Castillo, J. (2012). *Distribución online de contenidos audiovisuales: análisis de 3 modelos de negocio*. *Profesional de La Informacion, 21*(4), 385–390. https://doi.org/10.3145/epi.2012.jul.09

5.  Mareike,J.(2014) . *Is this TVIV? On Netflix, TVIII and binge-watching.*Independent Scholar, Ringstr 36, Berlin 12205, Germany.https://doi.org/10.1177%2F1461444814541523

6.  Praveen Seshadri, Miron Livny, and Raghu Ramakrishnan. 1996. *The Design and Implementation of a Sequence Database System.* In Proceedings of the 22th International Conference on Very Large Data Bases (VLDB '96). Morgan Kaufmann Publishers Inc., San Francisco, CA, USA, 99–110.
    https://dl.acm.org/doi/10.5555/645922.673634

7.  Ardizzone E., La Cascia M. (1997), *Automatic Video Database Indexing and Retrieval, Multimedia Tools and Applications* https://doi.org/10.1023/A:1009630331620

8.  Boll, S., Klas, W., & Sheth, A. P. (199*8). Overview on Using Metadata to Manage Multimedia Data. Multimedia Data Management:* Using Metadata to Integrate and Apply Digital Media, 1-23.
    https://corescholar.libraries.wright.edu/knoesis/809

9.  de Vries, A. P. (1998). *Mirror: Multimedia Query Processing in Extensible Databases. In D. Hiemstra, F. de Jong, & K. Netter (Eds.), Language Technology in Multimedia Information Retrieval: Proceedings of the Fourteenth Twente Workshop on Language Technology (TWLT14) (pp. 37-48). (Twente Workshop on Language Technology 14; No. 14)*. University of Twente. https://research.utwente.nl/en/publications/mirror-multimedia-query-processing-in-extensible-databases

10. de Vries, A. P., van der Veer, G. C., & Blanken, H. (1998). *Let's talk about it: dialogues with multimedia databases Database support for human activity.* Displays, 18(4), 215-220. [10.1016/S0141-9382(98)00023-7].
    https://doi.org/10.1016/S0141-9382(98)00023-7

11. *Brindha D, Jayaseelan R., Kadeswaran S.,(2021). Covid-19 Lockdown,entertainment, and paid ott video streaming platforms: A qualitative study of audience preferences . In Mass Communicator: International Journal of Communication Studies Volume 14 Issue 4*
    *https://www.indianjournals.com/ijor.aspx?target=ijor:mcomm&volume=14&issue=4&article=002*

12. *Stefan D., Nelson M.(2021) Integrating SQL Databases with Content-specific Search Engines. Proceedings of the 23rd VLDB Conference Athens, Greece, http://vldb.org/conf/1997/P528.PDF*

13. *Sundaravel E., Elangovan N. Emergence and future of Over-the-top (OTT) video services in India: an analytical research. https://www.journalbinet.com/uploads/2/1/0/0/21005390/50.02.08.2020_over-the-top__ott__video_services_in_india.pdf*

14. *Ninnet K.,Pichapop B.,Vichayuth(2020) S. Opportunities and Challenges of OTT (Over-The-Top) Services in Thailand. 2020: Proceedings of the 2020 ANPOR Annual Conference*
    https://journal.anpor.net/index.php/proceeding/article/view/164

15. *Philippe A., Hongjiang Z., Dragutin P.(1996) Content-based representation and retrieval of visual media: A state-of-the-art review. Multimed Tools Appl 3, 179–202*
    https://link.springer.com/article/10.1007/BF00393937

16. Samala N., Soumya S., Venkat Reddy Yasa,(2021) *Factors affecting consumers' willingness to subscribe to over-the-top (OTT) video streaming services in India, Technology in Society,* Volume 65, 2021, 101534
    https://www.sciencedirect.com/science/article/abs/pii/S0160791X21000099

17. Samriti, Divya , Sharma and  Priyank,(2020). *OTT- Existing Censorship Laws and Recommendations* . Published at ssrn.com
    https://papers.ssrn.com/sol3/papers.cfm?abstract_id=3735027

18. Bruce W. Herr; Weimao Ke; Elisha Hardy; Katy Borner,*Movies and Actors: Mapping the Internet Movie Database,Published in: 2007 11th International* Conference Information Visualization (IV '07)
    https://ieeexplore.ieee.org/abstract/document/4272022

19. Ebru,B.(2018)., *Methodology for the Regulation of Over-the-top (OTT) Services: The Need of A Multi-dimensional Perspective* . International Journal of Economics and Financial Issues; Mersin Vol. 8, Iss. 1
    https://search.proquest.com/openview/c327c4c8e7982e042326b58e2b193f3d/1?pq-origsite=gscholar&cbl=816338

20. Arif, G.(1995) .*Multimedia database management systems*;published at ACM Computing Surveys
    https://dl.acm.org/doi/abs/10.1145/234782.234798

21. Arjen , P. *CONTENT AND MULTIMEDIA DATABASE MANAGEMENT SYSTEMS*;published at Centre for Telematics and Information Technology (CTIT)
    doi=10.1.1.103.7992